\documentclass[twoside]{sfm}

\input{sf.def}

\begin{document}

\def\llm{{\sc LLmodels}}
\def\atl{{\sc ATLAS9}}
\def\aatl{{\sc ATLAS12}}
\def\starsp{{\sc STARSP}}
\def\aur{$\Theta$~Aur}
\def\logg{\log g}
\def\tauros{\tau_{\rm Ross}}
\def\kms{km\,s$^{-1}$}
\def\bz{$\langle B_{\rm z} \rangle$}
\def\degr{^\circ}
\def\aaps{A\&AS}
\def\aap{A\&A}
\def\apjs{ApJS}
\def\apj{ApJ}
\def\rmxaa{Rev. Mexicana Astron. Astrofis.}
\def\mnras{MNRAS}
\def\actaa{Acta Astron.}
\newcommand{\Tef}{T$_{\rm eff}$~}
\newcommand{\Vt}{$V_t$}
\newcommand{\CC}{$^{12}$C/$^{13}$C~}
\newcommand{\CDC}{$^{12}$C/$^{13}$C~}

\pagebreak

\thispagestyle{titlehead}

\setcounter{section}{0}
\setcounter{figure}{0}
\setcounter{table}{0}

\markboth{Drake et al.}{HgMn and PGa stars}

\titl{Establishing the link between HgMn and PGa stars}{Drake N. A.$^{1,2}$, Hubrig S.$^3$, Sch\"oller M.$^4$, 
Ilyin I.$^3$, Castelli F.$^5$, Pereira C. B.$^1$, Gonzalez J. F.$^6$}
{
$^1$Observat\'orio Nacional/MCTI, Rua Jos\'e Cristino 77, CEP 20921-400, 
     S\~ao Crist\'ov\~ao, Rio de Janeiro, RJ, Brazil, email: {\tt drake@on.br} \\
$^2$Sobolev Astronomical Institute, St.~Petersburg State University, Universitetski pr.~28, 
   198504, St.~Petersburg, Russia\\
$^3$Leibniz-Institut f\"ur Astrophysik, An der Sternwarte~16, 14482~Potdam, Germany\\
$^4$European Southern Observatory, Karl-Schwarzschild-Str.~2, 85748~Garching, Germany\\
$^5$Instituto Nazionale di Astrofisica, Osservatorio Astronomico di Trieste, 
     Via Tiepolo~11, I-34143~Trieste, Italy\\
$^6$Instituto de Ciencias Astronomicas, de la Tierra, y del Espacio (ICATE), 5400, San Juan, Argentina}

\abstre{
We discuss most recent spectroscopic and spectropolarimetric observations of the star
HD\,19400 representative of the group of PGa stars.
Our high-spectral-resolution study of abundances, line profile variability, and
the longitudinal magnetic field of HD\,19400 discloses a remarkable similarity between
this group and the group of HgMn stars. 
}

\baselineskip 12pt

\section{Introduction}
\label{sect:drake_intro}

The mid to late-main sequence B-type stars contain several groups of chemically peculiar stars,
among them classical  magnetic Bp stars, HgMn stars, and PGa stars.
The classical  magnetic peculiar Bp stars, which include Si and He-weak Bp stars,  are usually 
characterised by large overabundances of Fe-peak 
and rare-earth elements. They possess non-axisymmetric large-scale magnetic fields of up to a 
few kG and display on their surfaces chemical spots of different elements.  
On the other hand, the presence of weak magnetic fields on the surface of late-B type stars with 
HgMn peculiarity, the so-called HgMn stars,  has been controversial during the last two decades.
Our recent measurements of magnetic fields in HgMn stars with the moment technique, using spectral 
lines of several elements separately, revealed the presence of a weak longitudinal magnetic field, 
a quadratic magnetic field, and the crossover effect on the surface of several HgMn stars (Hubrig 
et al.\ 2012 \cite{drake_Hubrig2012}).

The aspect of inhomogeneous distribution of some chemical elements over the surface of HgMn stars 
was first discussed by Hubrig \& Mathys (1995 \cite{drake_Hubrig1995}). From a survey of HgMn stars in close 
spectroscopic binaries, it was 
suggested that some chemical elements might be inhomogeneously distributed on the surface, with, 
in particular, preferential concentration of Hg along the equator. 
Recent studies revealed that not only Hg, but also many other elements, most typically Ti, Cr, Fe, 
Mn, Sr, Y, and Pt, are concentrated in spots of diverse size, and different elements exhibit 
different abundance distributions across the stellar surface (e.g.,\ Hubrig et al.\ 2006 \cite{drake_Hubrig2006}; 
Briquet et al.\ 2010 \cite{drake_Briquet2010}; Korhonen et al.\ 2013 \cite{drake_Korhonen2013}).
Moreover, an evolution of the abundance spots of several elements at different time scales was 
discovered in 
two additional HgMn stars, HD\,11753 and AR\,Aur. 
Briquet et al.\ (2010 \cite{drake_Briquet2010}) and Korhonen et al.\ (2013 \cite{drake_Korhonen2013}) reported the 
presence of dynamical spot evolution over a couple of weeks for the SB1 system HD\,11753, 
while Hubrig et al.\ (2010 \cite{drake_Hubrig2010}) detected a secular element evolution in the double-lined eclipsing 
binary AR\,Aur.
Our recent spectroscopic and polarimetric studies of HgMn stars suggest the 
existence of intriguing correlations between the strength of the magnetic field, 
abundance anomalies, and binary properties (Hubrig et al.\ 2012 \cite{drake_Hubrig2012}). 
However, not much is known about the origin of anomalies in the hotter extension of 
the HgMn stars, the PGa stars, with rich P\,{\sc ii}, Mn\,{\sc ii}, Ga\,{\sc ii}, and 
Hg\,{\sc ii} spectra, and effective temperatures of about 14\,000\,K and higher. 
In this work, we present a spectroscopic study of the typical PGa star HD\,19400.

\section{The spectrum of  the PGa star HD\,19400}  

\begin{figure}[!t]     
\begin{center}
\vspace{-5mm}
 \includegraphics[width=0.42\textwidth, angle=0]{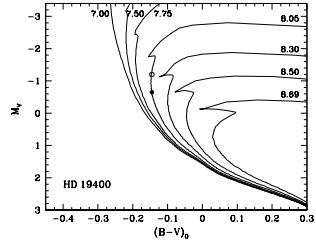}
\vspace{-9mm}
\caption[]{The position of HD\,19400 in the H-R diagram. The filled circle corresponds to 
the position obtained using the Hipparcos parallax, while the open circle  corresponds to 
spectroscopic $T_{\rm eff}$ and $\log g$ (Leone et al.\ 1997 \cite{drake_Leone1997}).
}
\label{fig:drake_hr}
\end{center}
\end{figure}

We recently downloaded two 
high-resolution, high S/N HARPSpol spectra of the PGa star HD\,19400 from the ESO archive to 
study the presence of a magnetic field and an inhomogeneous distribution of overabundant elements on the 
stellar surface.
A careful inspection of the polarimetric spectra of HD\,19400 acquired on two consecutive nights 
in 2011 revealed the presence of anomalous flat-bottom line profiles belonging to overabundant 
elements, reminiscent of profile shapes observed in numerous HgMn stars. 
Magnetic field measurements using the moment technique introduced by 
Mathys (e.g.\ 1991 \cite{drake_Mathys1991}) disclosed the presence of a weak mean longitudinal  
magnetic field of the order of a few tens of Gauss, similar to weak fields 
detected in a number of HgMn stars (e.g., Mathys \& Hubrig 1995 \cite{drake_Mathys1995}; Hubrig et al.\ 2012 
\cite{drake_Hubrig2012}).
To date, HD\,19400 is the only PGa star for which a search for the presence of a magnetic
field and an inhomogeneous element distribution has been carried out. The position of 
HD\,19400 in the H-R diagram is presented in Fig.~\ref{fig:drake_hr}.

\begin{figure}[!t] 
\vspace{-5mm}
\begin{center}
\includegraphics[width=0.42\textwidth, angle=0]{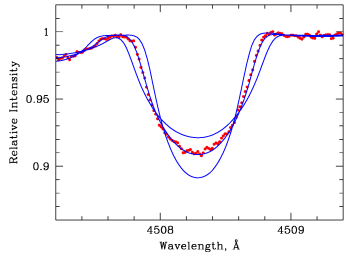}
\vspace{-4mm}
\caption[]{
Determination of the $v\,\sin\,i$ value for HD\,19400. Solid blue lines 
correspond to profiles calculated with $v\,\sin\,i$ = 25, 30, and 
35 km\,s$^{-1}$. The observed line profile is shown by red dots. 
}
\label{fig:drake_rot}
\end{center}
\end{figure}

As shown in Fig.~\ref{fig:drake_rot}, the $v\,\sin\,i$ value of 30~km\,s$^{-1}$ was 
estimated from fitting a synthetic spectrum to the observed profile of the Fe\,{\sc ii} 
$\lambda$4508.3~\AA\ line, which has a low Land\'e factor. 

\begin{figure}[!t] 
\vspace{-11mm}
\begin{center}
\includegraphics[height=0.43\textwidth, angle=0]{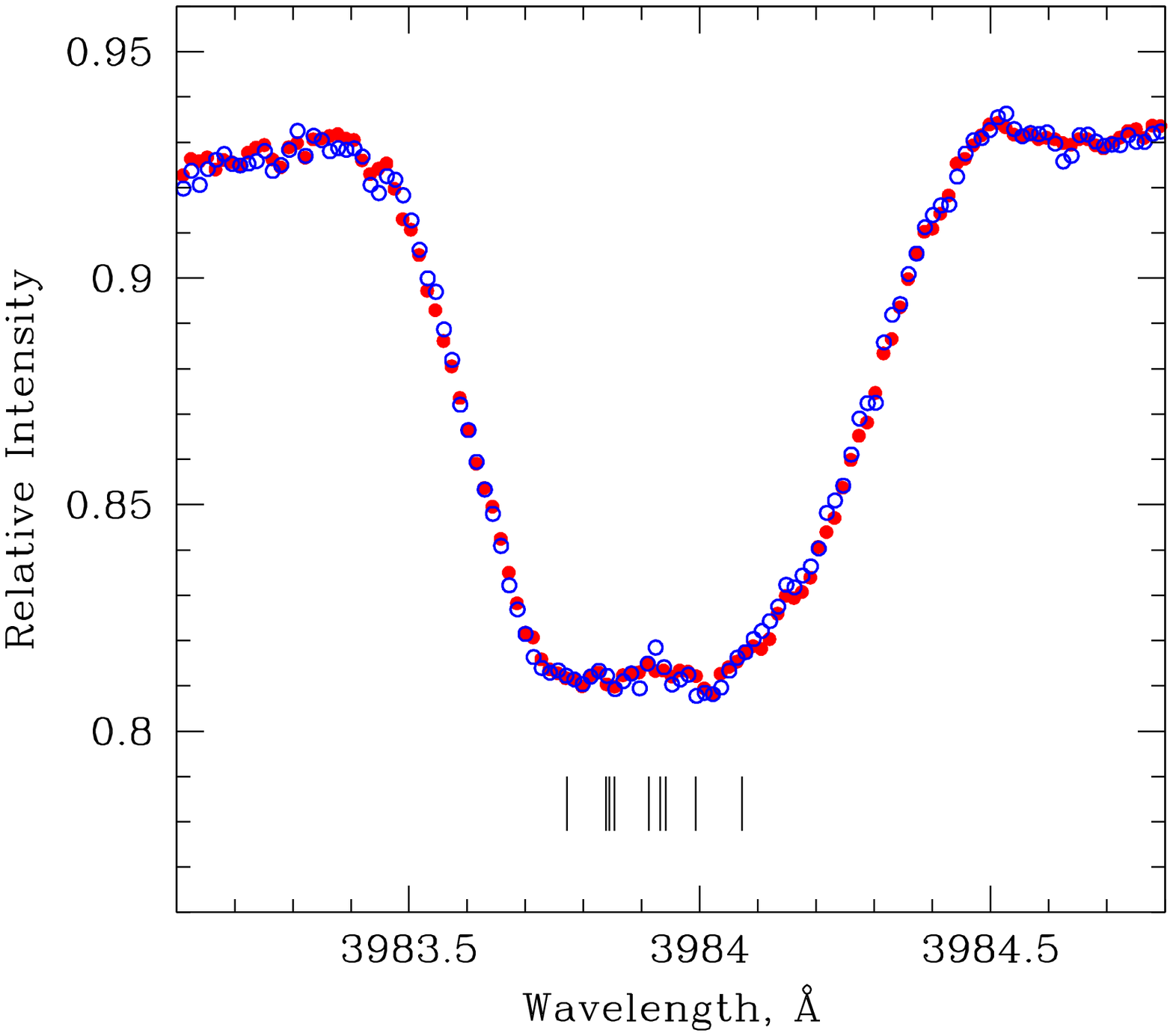}
\includegraphics[height=0.51\textwidth, angle=0]{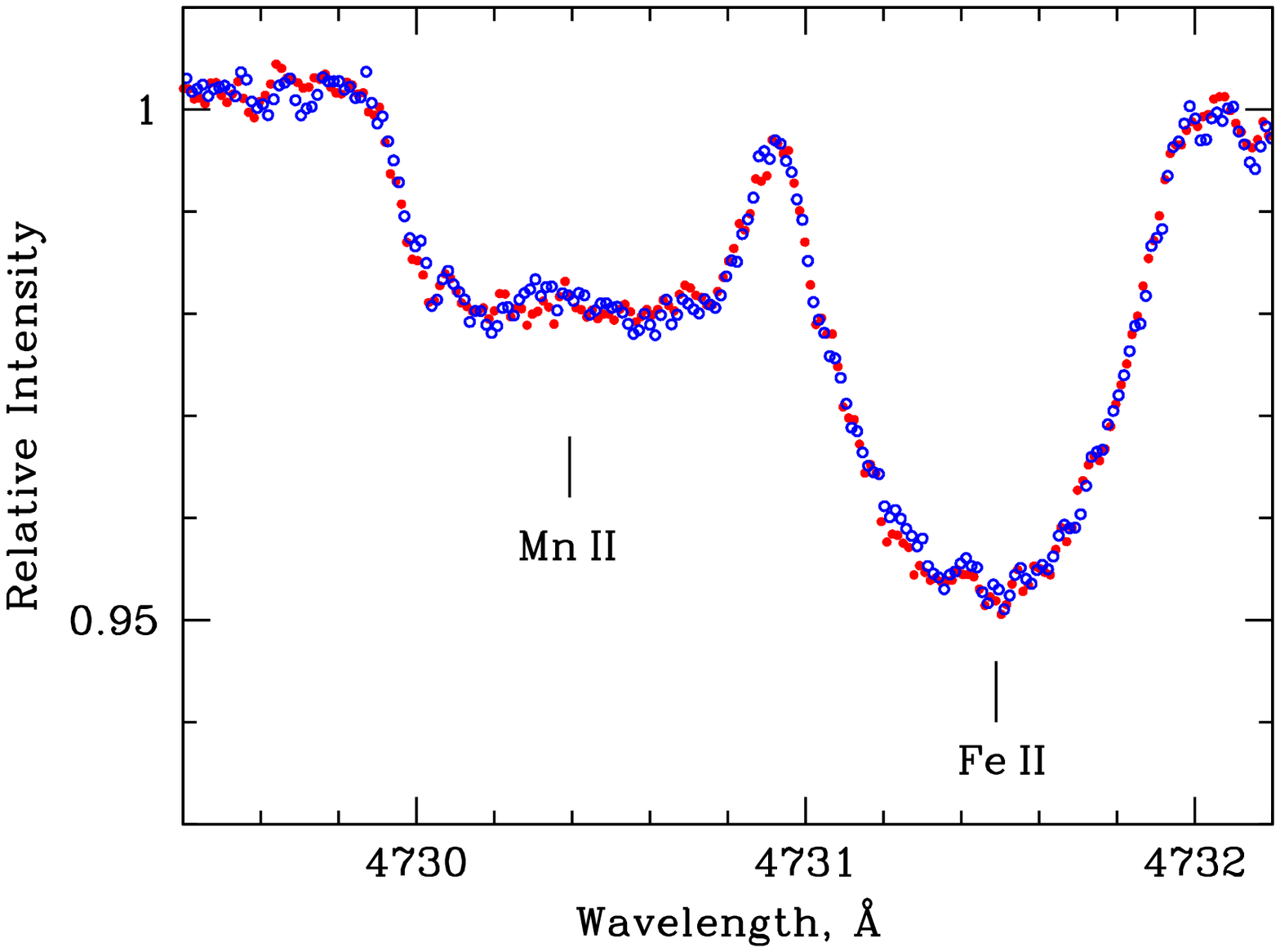}
\vspace{-9mm}
\caption[]{
Observations of line profiles belonging to three different elements 
are highlighted for different nights by open and filled circles.
{\em Left:} The Hg\,{\sc ii} $\lambda$3984~\AA\ line together with 
the indicated isotopic/hyperfine structure. 
{\em Right:} Profiles of  Mn\,{\sc ii} and Fe\,{\sc ii} lines in the 
spectral region around 4730.5\,\AA{}.
}
\label{fig:drake_flat}
\end{center}
\end{figure}

The HARPSpol spectra observed on two consecutive nights on December 15 and December 16,
2011, show tiny variations in the line shapes.
In Fig.~\ref{fig:drake_flat} we present these weak variations in several line 
profiles. Moreover, Fig.~\ref{fig:drake_flat} clearly shows that the shape of the 
line profiles belonging to Hg\,{\sc ii} and Mn\,{\sc ii}  strongly deviates from the rotationally 
broadened Fe\,{\sc ii} profiles, indicating an inhomogeneous distribution of at 
least these elements on the stellar surface.

\section{Synthesis of line profiles of Hg\,{\sc ii} and Mn\,{\sc ii} taking into account 
the isotopic and hyperfine structure} 

\begin{figure}[!t]
\begin{center}
\includegraphics[height=0.45\textwidth, angle=0]{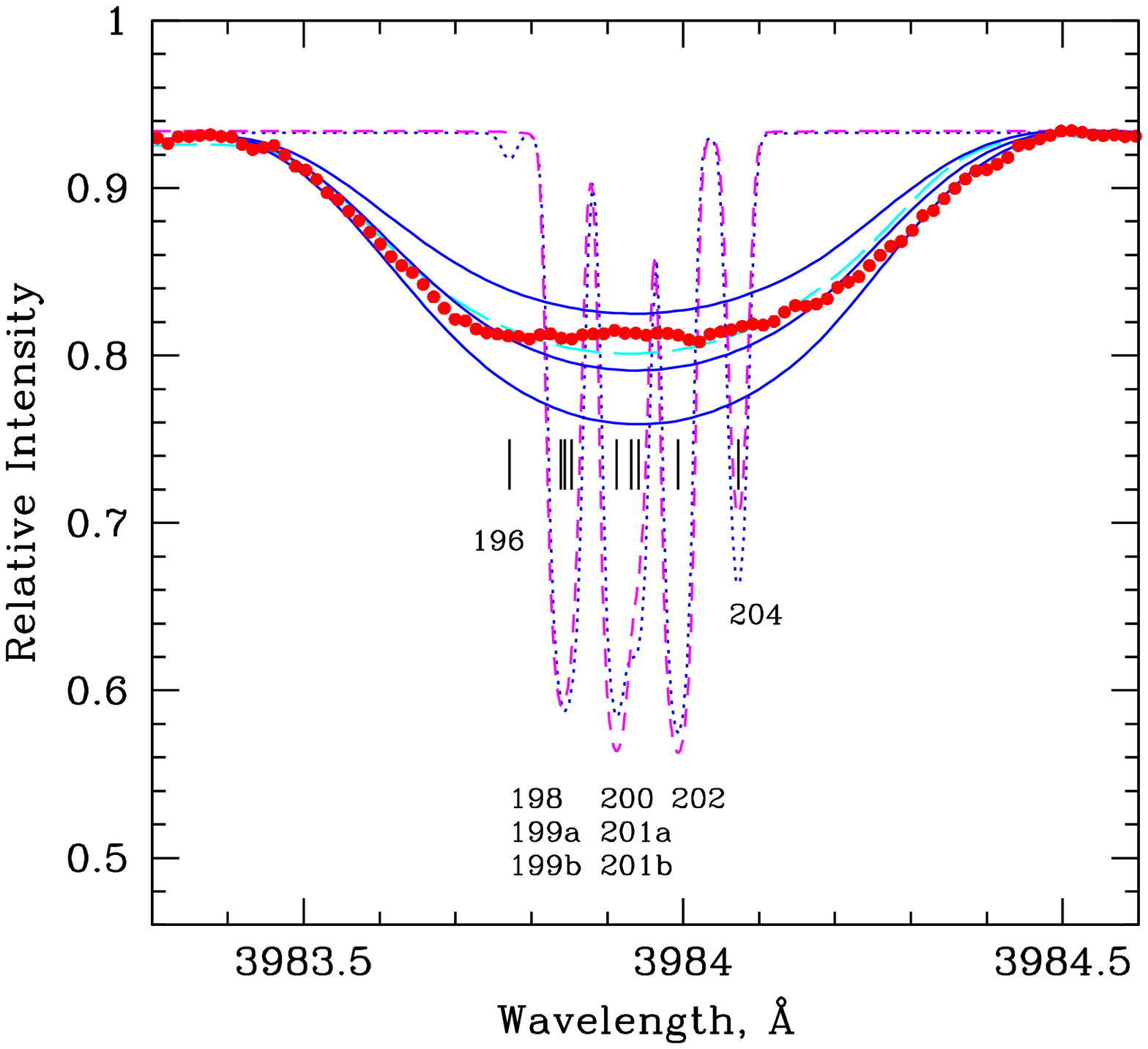}
\includegraphics[height=0.45\textwidth, angle=0]{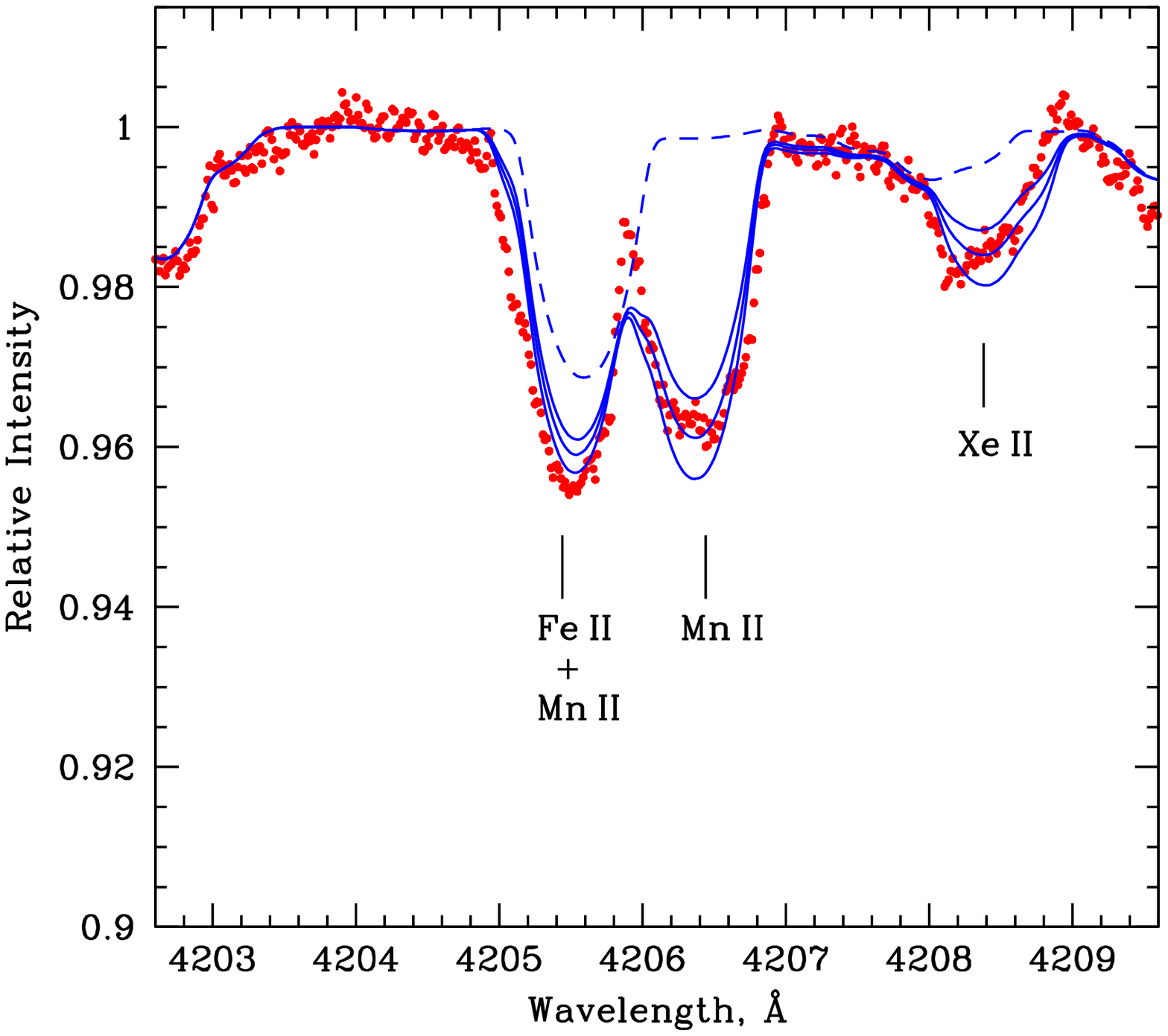}
\vspace{-5mm}
\caption[]{
{\em Left:} Synthetic spectrum in the region around the Hg\,{\sc ii} $\lambda$3984~\AA\ line.
 {\em Right:} Synthetic spectrum in the region around the Mn\,{\sc ii} $\lambda$4206~\AA\ line.
Red points correspond to observations, while solid blue lines indicate the synthetic 
spectra calculated for different abundances of Hg (for $\log \epsilon$(Hg) = 5.7, 6.0, 6.3) 
and of Mn (for $\log \epsilon$(Mn) = 7.20, 7.25, 7.30). 
The blue dotted line in the figure on the left side represents 
a terrestrial isotopic mixture (Dolk et al.\ 2003 \cite{drake_Dolk2003}) 
without broadening, while for the magenta short dash line we 
assume the isotopic composition obtained by Castelli \& Hubrig (2004 
\cite{drake_Castelli2004}) for the HgMn star HD\,175640.
The long dash cyan line represents the Hg\,{\sc ii} line profile 
calculated without consideration of hyperfine structure  and isotopic composition.  
The short dashed blue line in the figure on the right side presents the synthetic spectrum calculated 
with solar abundances of Mn and Xe (Anders \& Grevesse 1989 \cite{drake_Anders1989})
}
\label{fig:drake_hfs}
\end{center}
\end{figure}

In Fig.~\ref{fig:drake_hfs} we present synthetic spectra in the spectral regions 
containing the Hg\,{\sc ii} $\lambda$3984~\AA\ and Mn\,{\sc ii} $\lambda$4206~\AA\ lines.
Synthetic line profiles have been calculated using Kurucz models, the current version of
the {\sc moog} code (Sneden 1973 \cite{drake_Sneden1973}), atmospheric parameters of 
Leone et al.\ (1997 \cite{drake_Leone1997}), and $v_{\rm t} = 0.2$~km\,s$^{-1}$. 
Obviously, neither the impact of isotopic/hyperfine structure of Hg on the profile shape, nor the 
hyperfine structure of the Mn\,{\sc ii} lines can produce the observed  flat-bottom line profiles.

\section{Appearance of line profiles in typical HgMn stars} 

\begin{figure}[!t]
\begin{center}
\includegraphics[height=0.40\textwidth, angle=0]{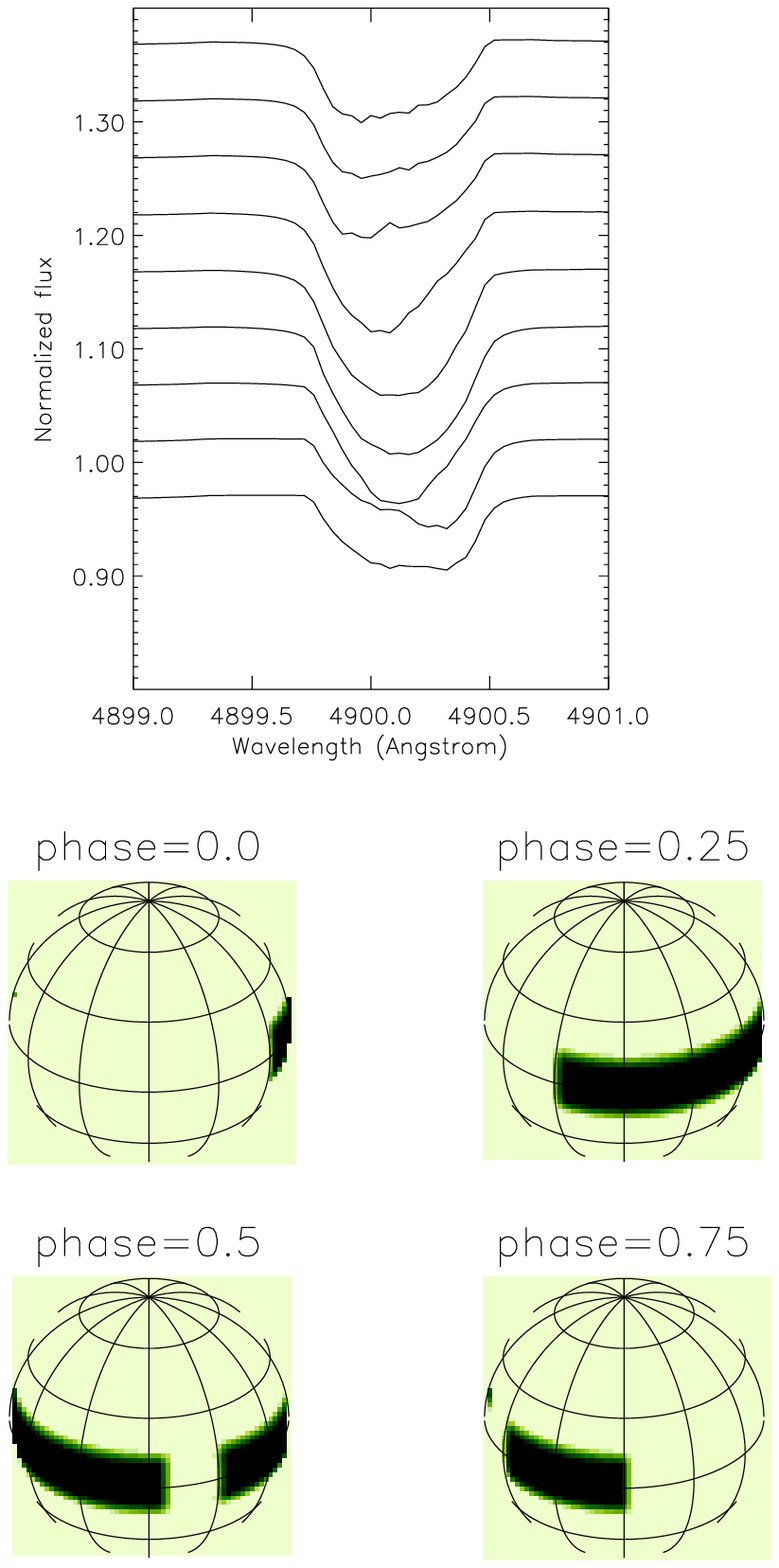}
\includegraphics[height=0.40\textwidth, angle=0]{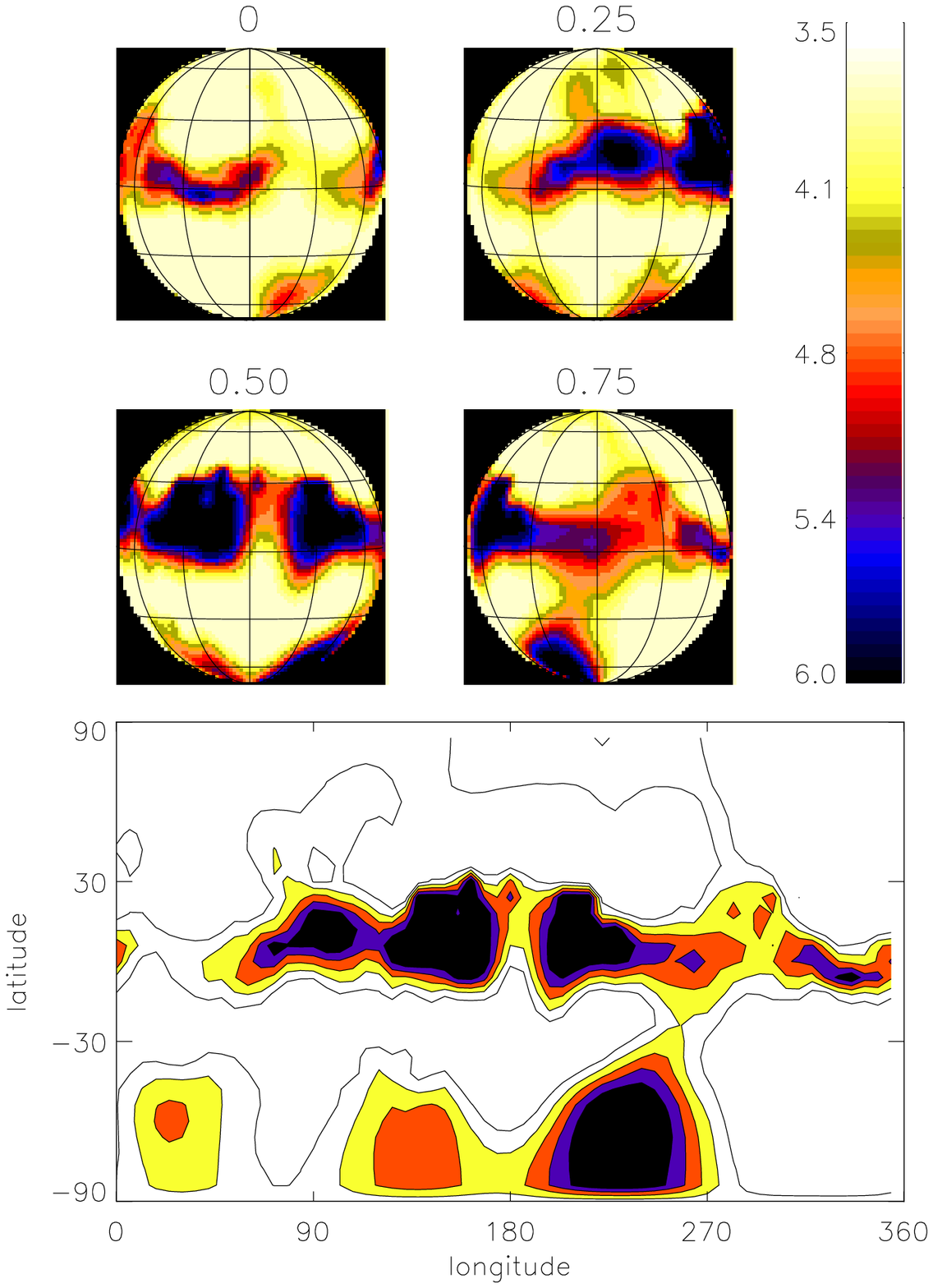}
\vspace{-1mm}
\caption[]{
{\em Left:} Observed Y\,{\sc ii} $\lambda$4900~\AA\ line variations in 
UVES spectra of the eclipsing binary AR\,Aur with a HgMn primary. 
Using a direct Doppler imaging method, the flat-bottom Y\,{\sc ii} profiles can be 
explained by the presence of a Y\,{\sc ii} fractured equatorial ring 
(Hubrig et al.\ 2006 \cite{drake_Hubrig2006}). 
{\em Right:} A surface Y\,{\sc ii} abundance  distribution similar to a fractured 
equatorial ring was detected using the Doppler imaging 
inversion code iAbu (Savanov et al.\ 2009 \cite{drake_Savanov2009}).
}
\label{fig:drake_di}
\end{center}
\end{figure}

Similar flat-bottom line profiles belonging to overabundant elements are frequently observed in 
typical HgMn stars. 
In Fig.~\ref{fig:drake_di}, we present UVES spectra of the eclipsing binary AR\,Aur 
with a HgMn primary, obtained a few years ago. 

\section{Conclusions}

Due to a small number of detailed high-resolution, high S/N spectropolarimetric 
studies of PGa stars, the link between the chemically peculiar groups among the 
mid to late main sequence B-type stars is not understood yet. 
Apart from our recent study of HD\,19400, no other information on spectral 
variability and the presence of magnetic fields in PGa stars currently
exists. 
Our study of the PGa star HD\,19400 indicates a spectral behaviour very 
similar to that of HgMn stars showing the presence of chemical spots and 
weak magnetic fields.
Future high S/N, high-resolution spectropolarimetric observations of PGa 
stars are important to firmly establish the link between PGa and HgMn stars.

\vspace{-2mm}
{\it Acknowledgements.} N.A.D. acknowledges support of the PCI/MCTI, Brazil, under the 
Project 302350/2013-6 and the Saint Petersburg State University, Russia, under the Project 6.38.73.2011.

\end{document}